\documentclass[english, 12pt]{article}
\usepackage[english]{babel,varioref}
\usepackage[ansinew]{inputenc}
\usepackage[fleqn,reqno]{amsmath}
\usepackage{paralist}
\usepackage{natbib}
\usepackage{graphics}
\usepackage{epic,eepic}

\bibpunct[: ]{(}{)}{,}{a}{}{,}

 \setdefaultenum{(i)}{a)}{(1)}{(A)}

\usepackage{setspace}

\begin{document}

\title{On the Logic (plus some history and philosophy) of Statistical Tests and Scientific Investigation}

\author{Uwe Saint-Mont, Nordhausen University of Applied Sciences}

\maketitle

\section*{Abstract}
Every scientific endeavour consists of (at least) two components: A hypothesis on the one hand and data on the other. There is always a more or less abstract level - some theory, a set of concepts, certain relations of ideas - and a concrete level, i.e., empirical evidence, experiments or some observations which constitute matters of fact.

The focus of this contribution is on elementary models connecting both levels that have been very popular in psychological research - statistical tests. Going from simple to complex we will examine four paradigms of statistical testing (Fisher, Likelihood, Bayes, Neyman \& Pearson) and an elegant contemporary treatment.

In a nutshell, testing is an easy problem that has a straightforward mathematical solution. However, it is rather surprising that the statistical mainstream has pursued a different line of argument. The application of the latter theory in psychology and other fields has brought some progress but has also impaired scientific thinking.
\footnote{Key words and phrases: Statistical testing; tests of hypotheses; scientific thinking; modes of inference; history of statistics; foundational issues}

\newpage

\section{Fisher: One Hypothesis}

\begin{quote}
Every experiment\label{exps1} may be said to exist only to give the facts a
chance of disproving the null hypothesis. (\citet{fi35}, p. 16)
\end{quote}

The simplest and oldest formal model is Fisher's test of significance. There is just one distribution, called the ``hypothesis'' $H$ (or $P_H$), and a sample from this population. Formally, the random variables $X,X_1,\ldots, X_n$ are iid, $X \sim P_H$, and $X_i=x_i$ are the observations subsequently encountered ($i=1,\ldots,n$). Thus ${\bf x}=(x_1,\ldots,x_n)$ is the vector of data at hand.

In the continuous case, $X$ has a density $f(x)$, whereas in the discrete case, $X$ assumes values $x_1,x_2,\ldots$ with corresponding probabilities $p_1,p_2,\ldots$ Information theory often restricts attention to random variables assuming values in a finite alphabet ${\cal X} = \{x_1,\ldots,x_k\}$. In the following, in order to keep technical issues to a minimum, a random variable will be discrete if not otherwise stated.

Given this setting, suppose one observes a single $x$ for which $p=P_H(X=x)=0$. That is, this observation should not have occurred since the hypothesis does not account for it. It is simply impossible to see $x$ if $P_H$ is the case. In other words, this concrete observation $x$ falsifies the hypothesis $P_H$, it is a counterexample to the law $P_H$. In philosophical jargon, this is a strict, logical conclusion (the modus tollens). One concludes without any doubt, although formalized by the probability statement ($p=0$), that the hypothesis in question is not the case.

Now, what if $p$ is ``small''? Obviously, no matter how small the probability, as long as $p>0$, the observation $x$ is possible and we cannot infer with rigour that $X \sim P_H$ did not produce it. All one may say is that

\begin{quote} {\it Either} an exceptionally rare chance has occurred, {\it or}
the theory [hypothesis] of random distribution is not
true. (\citet{fi73}, p. 42, emphasis in the original)
\end{quote}

Of course, such an ``inductive statistic'' (IS) argument is much weaker than ``deductive nomological'' (DN) conclusions, like the one considered before, and consequently a lot of discussion has been spawned by Fisher's dichotomy (see, e.g., \citet{sa89, fe01}). More important to our reasoning is the observation that no general statistical theory evolved from Fisher's dichotomy. Here are two reasons why: First, if $X$ assumes $k$ distinct values $x_1,\ldots,x_k$ with probabilities $p_1,\ldots,p_k$, ``small'' crucially depends on the number of possible observations $k$. (A probability of $p=1/100$ is small if $k=10$, however the same probability is rather large if $k=10^6$ or $k=10^{20}$, say.) Second, if $P_H$ is the uniform distribution, we have no reason whatsoever to discard the hypothesis no matter which value occurs. Each and every $x_j$ is equally (un)likely, but possible nonetheless. If $P_H$ has a geometric distribution, i.e., if $X$ assumes the natural number $j$ $(j \ge 1)$ with probability  $2^{-j}$ it would be very difficult to tell beyond which number $j_0$ the probability $p_{j_0}$ could be said to be ``small''. Finally, for any continuous distribution, in particular the standard normal, we have $P(X=x)=0$ for every $x$. However, since some realization must show up, some $x$ will occur nevertheless.

Perhaps for reasons such as these, Fisher came up with a more sophisticated idea. Typically, most values observed are rather ``moderate'', and only a few are ``extreme'' outliers (very large or very small). Suppose large values of $X$ are suspicious. Then, having encountered $x$, it is straightforward to calculate $p=P_H(X \ge x)$, the probability of observing a value at least as large as $x$. If this so-called $p$-value is small, we have reason to reject the hypothesis. Of course, if small values of $X$ are suspicious, it is $P_H(X \le x)$ that should be considered, and in the case of outliers to the left and to the right of the origin, $P(X\ge |x|)$ is of interest. Thus we have a general rule: Calculate the probability of the value observed and of all ``more extreme'' events. This kind of evaluation may be crude, but it is also a straightforward way to quantify the evidence in the data $x$ about the hypothesis $P_H$.

 In the earliest test of this kind recorded, \citet{ar10} looked at London births. His hypothesis was that it is equally likely to have a boy or a girl. (Why should one of the sexes be preferred?) Considering a moderate number $n$ of years altogether, it would not be astonishing if the boys outnumbered the girls in about $n/2$ years, but by the laws of probability it would also not be surprising if there were more girls in perhaps 20 out of 30 years. However, it would be very surprising if, over a longer period of time, one sex outnumbered the other permanently. As a matter of fact, Arbuthnot checked $n=82$ successive years and learned that in each and every year more boys than girls were born. If $P_H(boy)=P_H(girl)=1/2$, the probability of the event ``boys always outnumbering girls'' happening by chance is $2^{-82}$. Thus he concluded that some force ``made'' more boys than girls.

 Suppose Arbuthnot had found 80 years with more boys than girls. Then Fisher's advice is to calculate
 \begin{eqnarray}\label{arbu} p &=& P_H(X \ge 80)=P_H(X=82)+P_H(X=81)+P_H(X=80)
 \\\nonumber
 &=& \binom{82}{82} 2^{-82} + \binom{82}{81} 2^{-82} + \binom{82}{80} 2^{-82} = \frac{1+82+3321}{2^{82}}= \frac{3404}{2^{82}} \approx 7 \cdot 10^{-22}.
 \end{eqnarray}

 Since all probabilities sum up to one, this seems to be a small value, and thus a remarkable result.  \citet{fi29}, p. 191, writes:
 \begin{quote}
 It is a common practice to judge a result significant, if it is of such a
  magnitude that it would have been produced by chance not more frequently than
  once in twenty trials. This is an arbitrary, but convenient, level of significance
  for the practical investigator $[\ldots]$
 \end{quote}
Today, the standard levels of significance are $5\%, 1\%,$ and $0.1\%$. Although, ``surely God loves the 0.06 nearly as much as the 0.05?'' \citep{ro89}

\subsection{Objections}\label{object}

Despite the above rather natural derivation, problems with $p$-values and their proper interpretation turned out to be almost endless:

 The smaller the $p$-value, the larger the evidence against some hypothesis $H$, an idea already stated explicitly in \citet{be42}. Thus one should be able to compare $p$-values or combine $p$-values of different studies. Unfortunately, if two experiments produce the same $p$-value, they do not provide the same amount of evidence, since other factors, in particular the total number of observations $n$, also play a considerable role \citep[19]{co66}.

\citet{jo86}, p. 496, elaborates: ``Thus,
  as Jeffreys explained in 1939, if the sample is very large, the level of significance
  $P$ tends to exaggerate the evidence against the null hypothesis, i.e. $P$ tends
  to be smaller than it ought to be. But in practice, if the sample is very
  large, a good orthodox statistician will `deflate' intuitively the level of
  significance $P$ accordingly.''
  \citet{mc89} comments on this: ``This is very
  likely true, but it is an inadequate base for presenting the $p$ value
  approach to scientists.''

The best one can do seems to be rules of thumb.
For example, \citet{ef01}, p. 212, consider the normal distribution and sample size $n$ in order to translate $p$-values into evidence.
However, \citet{ro86} demonstrates that contradictory statements are possible: ``A given $P$-value in a large trial is usually stronger
evidence that the treatments really differ than the same $P$-value in a small
trial of the same treatments would be'' (\citet{pe76}, p. 593). But also ``The
rejection of the null hypothesis when the number of cases is small speaks for a
more dramatic effect $[\ldots]$ if the $p$-value is the same, the probability of
committing a Type I error remains the same. Thus one can be more confident with
a small $N$ than a large $N$'' (\citet{ba70}, p. 241) is a reasonable line of argument.

In a nutshell, it is very difficult to interpret and combine $p$-values in a logically satisfactory way (see
\citet{gr16, hu08} for recent overviews). \citet{sc96}, p. 126, also collects common ideas, in particular,
\begin{quote}
If my findings are not
    significant, then I know that they probably just occurred by chance and that
    the true difference is probably zero. If the result is significant, then I
    know I have a reliable finding. The $p$ values from the significance test
    tell me whether the relationship in my data are large enough to be important
    or not. I can also determine from the $p$ value what the chances are that
    these findings would replicate if I conducted a new study
\end{quote}
and then concludes that ``every one of these thoughts about the benefits of significance
testing is false.''
The most devastating point, however, seems to be the following consideration.

\subsection{The Observed and the Unobserved}\label{observe}

The distinction between the observed and the unobserved is fundamental to science. Science is built on facts, not speculation. Why have eminent statisticians confounded these two areas?

\begin{quote}
    It is not difficult to see how `Student' and Fisher found themselves
    defending the use of the $P$ integral. For if one accepts that it is
    possible to test a null hypothesis without specifying an alternative, and
    that the test must be based on the value of a test statistic in conjunction
    with its known sampling distribution on the null hypothesis, then the
    integral of the distribution between specified limits is the only
    measure which is invariant to transformation of the statistic. It follows
    that one is virtually forced to consider the area between the realized value
    of the statistic and a boundary as the rejection area - the $P$ integral, in
    fact. (\citet{ed92}, p. 178)
\end{quote}

 In other words, although the last paragraph can be interpreted as an invariance argument in favour of $p$-values,\footnote{Even if the measuring process is rather arbitrary, and only the ordering of the values recorded corresponds to something real, the $p$-value makes sense, since $P(X\ge x)=P(f(X) \ge f(x))$ for any monotone transformation $f$.} Fisher, considering a single hypothesis, simply had no other choice but to calculate $P$ integrals such as $(\ref{arbu})$.
 He knew that this way to proceed was not really sound:
  \begin{quote} Objection has sometimes been made that
  the method of calculating confidence limits by setting an assigned value such
  as $1\%$ on the frequency of observing 3 or
  less $[\ldots]$ is unrealistic treating values less than 3, which have not
  been observed, in exactly the same manner as 3, which is the one that has been
  observed. This feature is indeed not very defensible save as an
  approximation. (\citet{fi73}, p. 71)\end{quote}

  However, a rather straightforward example illustrates that even the roundabout idea of ``approximation'' is difficult to defend. Suppose $P_H(X
<x)=0.01$ and $P_H(X=x)=0.02$, small values of $X$ being suspicious. If $x$ is observed, the one-sided test may reject $P_H$ since $P_H(X \le x) =0.03$. Now look at the (modified) hypothesis $K$ where $P_K(X=x)=0.02$, but
  $P_K(X<x)=0.4$. In this case $P_K(X \le x)=0.42$ and no test would reject $K$. Yet the probability of the observed value $x$ is the same for both hypotheses! The conclusion differs tremendously just because of values that were {\it not} observed:

  \begin{quote} An hypothesis that may
  be true is rejected because it has failed to predict observable
  results that have not occurred. This seems a remarkable procedure.
  On the face of it, the evidence might more reasonably be taken as
  evidence for the hypothesis, not against it. (\citet{je39}, p. 316) \end{quote}

 Altogether, Fisher's paradigm seems to be too coarse. What is needed are more elaborated models, able to distinguish between observed and merely possible values, and explicitly formalizing other relevant aspects, such as the probability of committing an error or the strength of some effect.

\section{Two hypotheses}

 In order to keep things as simple as possible, \citet{pe38}, p. 242, proposed the following move:
  \begin{quote} $[\ldots]$ the only valid reason for rejecting
  a statistical hypothesis is that some alternative hypothesis explains the
  observed events with a greater degree of probability.\footnote{As early as 1926, Gosset wrote to E.S. Pearson: ``$[\ldots]$ if there is any alternative hypothesis $[\ldots]$ you will be much more inclined to consider that
  the original hypothesis is not true $[\ldots]$'' (See \citet{ro97}, p. 68,
and the discussion in \citet{ho90}, pp. 76.)}
\end{quote}

Given (at least) two hypotheses $H$ and $K$, it is of fundamental importance to understand that there are {\it two completely different ways to generalize} Fisher's approach. Either one sticks with integrals, which is the main feature of the Neyman-Pearson theory, or one directly compares $P_H(x)$ with $P_K(x)$. We will start with the latter idea:

\subsection{Likelihood Ratio Tests}

Given two hypotheses, it is perhaps most obvious to study the ratio $P_K(x) / P_H(x)$. In particular, since
  ``$\ldots$ a proper measure of strength of evidence should not depend on
  probabilities of unobserved values'' (\citet{ro97}, p. 69).
Obviously, a ratio larger than one is evidence in favour of $K$, and a ratio that is smaller than one provides evidence in favour of $H$. 

With successive observations $x_1,x_2,\ldots$ evidence for (and against) some hypothesis should build up. Mathematically, it is straightforward to consider the likelihood ratio, i.e., the product
\begin{equation}\label{like1}
r_n=r_n(x_1,\ldots,x_n)=\prod_{i=1}^n \frac{P_K(x_i)}{P_H(x_i)}.
\end{equation}
With every observation, the odds change in favour of one of the hypotheses (and thus, simultaneously, against the other). Let $P_{X^n}$ be the empirical distribution of a sample of size $n$. Due to the law of large numbers, $P_{X^n}(x) \rightarrow P_H(x)$ for every $x \in {\cal X}$ almost surely, if $P_H$ is the true distribution. This basic result almost immediately implies the likelihood convergence theorem: That is, (\ref{like1}) converges almost surely to zero if $H$ is true, and to $+\infty$ if $K$ is true. (See \citet{ro97}, p. 32, for discrete probability distributions and \citet{ch97}, p. 257, for densities.)


It thus seems to be justified to decide in favour of $K$ if the likelihood ratio exceeds some pre-assigned threshold $s$ $(s>1)$. As \citet{ro70} showed, if $H$ is correct, the probability that the ratio at one point of time exceeds $s$ is just $1/s$. Formally:
 $$
P \left(\prod_{i=1}^n \frac{P_K(X_i)}{P_H(X_i)} \ge s \;\; \mbox{for some} \;\; n=1,2,\ldots \right) \le \frac{1}{s}
 $$

Notice that even ``if an unscrupulous researcher sets out deliberately to find evidence supporting his favourite hypothesis [$K$] over his rival's
[$H$], which happens to be correct, by a factor of at least [$s$], then the chances are good that he will be eternally frustrated'' (\citet{ro97}, p. 7).

Since the normal distribution is particularly important, \citet{ro97}, p. 52, considers it in much detail and finds that $s=8$ and $s=16$, or $s=1/8=0.125$ and $s=1/16=0.0625$, respectively, are reasonable choices. For more details see \citet{ro00}, \citet{go88a}, and \citet{bo14}, p. 194, who reproduces Jeffreys' rule of thumb: $r_n>1$ supports $K$, $1 > r_n > 0.3$ supports $H$, ``but not worth more than a bare comment.'' However, the evidence in favour of $H$ (and thus, equivalently, against $K$) is

\begin{tabular}{|lll|lll|}
  \hline
  substantial & if & $0.3 > r_n > 0.1$ & very strong & if & $0.03 > r_n > 0.01$ \\
  strong & if & $0.1 > r_n > 0.03$  &  decisive & if & $0.01 > r_n$  \\
  \hline
\end{tabular}


\subsection{Bayesian Tests}\label{bayes}

The likelihood ratio may serve as the core piece of a Bayesian analysis.
 To this end let $\pi_H$ be the prior probability of the first hypothesis, and $\pi_K=1-\pi_H$ the prior probability of the second. Having observed ${\bf x}=(x_1,\ldots,x_n)$, Bayes' theorem states that the odds ratio of the posterior probabilities of the hypotheses is
\begin{equation}\label{like}
\frac{\pi(K|x_1,\ldots,x_n)}{\pi(H|x_1,\ldots,x_n)}= r_n(x_1,\ldots,x_n)  \cdot \frac{\pi_K}{\pi_H}= \prod_{i=1}^n \frac{P_K(x_i)}{P_H(x_i)}  \cdot \frac{\pi_K}{\pi_H}.
\end{equation}
  If $0<\pi_H<1$, i.e., if both hypotheses are considered possible at the beginning, there are convergence results of a very general nature that guarantee that the true hypothesis will be found almost surely (e.g., \cite{wa03, wa04}).

Moreover, it is possible to emulate Fisher's idea of a {\it single} explicit hypothesis. (For an example, see \citet{bo14}, pp. 197.)


\subsection{Neyman and Pearson}

Mathematicians J. Neyman and E.S. Pearson also improved upon Fisher's initial idea. In theory as well as in applications, their line of reasoning has become standard. Like Fisher, they used integrals, i.e., probabilities like $P(X \ge x)$. However, in order to avoid confounding the observed with the unobserved, they insisted that such probabilities be computed in advance, i.e., {\it before} recording empirical data.

Their paradigm situation is as follows: Denote by $N(\mu,\sigma)$ the normal distribution with expected value $\mu$ and standard deviation $\sigma$. Let $P_H \sim N(\mu_H,\sigma)$, $P_K \sim N(\mu_K,\sigma)$, and suppose without loss of generality that the absolute effect size $\eta = \mu_K-\mu_H$ is non-negative. Since for both hypothesis and each $x$ the densities $\varphi_H(x)$ and $\varphi_K(x)$ are positive, we can never be sure which hypothesis is the case. All we can do is try to minimize the error of the first kind (a decision in favour of $K$, although $H$ is true) and the error of the second kind (a decision in favour of $H$, although $K$ is true).

Given population $H$ or $K$, the mean ${\bar X}_n = \sum X_j /n$ of the observations is also normally distributed with parameters ${\mu}^{'}$, the correct hypothesis' expected value, and standard deviation $\sigma / \sqrt{n}$. (Thus, the larger the sample, the smaller the mean's standard deviation.) A rather straightforward treatment of this situation would look for the point $m$ where $\varphi_H(x)=\varphi_K(x)$ which, due to symmetry, is just $m=(\mu_H+\mu_K)/2$, and decide in favour of $H$ if $x < m$, and in favour of $K$ if $x \ge m$. This leads to the total probability of error
\begin{equation}\label{total}P_e(n) = \alpha_n + \beta_n = P({{\bar X}_n \ge m}|H)+P({\bar X}_n < m |K)
\end{equation}
which can be made arbitrarily small with growing $n$, for any fixed $\eta =\mu_K- \mu_H>0$.

However, perhaps since the errors of the first and of the second kind have different consequences, Neyman and Pearson decided to treat the null hypothesis $H$ (typically representing ``no effect'') and the alternative $K$ (representing a substantial effect) {\it asymmetrically}.
With $n$ and the effect size $\eta$ thus given,
\citet{ne33}, pp. 79, advised as follows:
\begin{quote}
From the point of
view of mathematical theory all that we can do is to show how the risk of the
errors $[\alpha, \beta]$ may be controlled and minimized. The use of these
statistical tools in any given case, in determining just how the balance
[between the two kinds of errors] should be struck, must be left to the
investigator.
\end{quote}
They also fixed $\alpha$ (i.e, the level of error of the first kind, meaning that an effect is detected although there is none). Now they could look for the optimum decision procedure, minimizing $\beta$, which they determined in \citet{ne33}.

Knowing the best test, one can also control for the errors (e.g., by fixing $\alpha$ to $0.01$, and assuming $\beta=0.3$, say), and set out to detect an effect of a certain size $\eta$ with the minimum number of observations $n$ necessary.
\citet{pe55}, p. 207, explains:\label{espearson1} \begin{quote} The appropriate test is
one which, while involving (through the choice of its significance level
$[\alpha]$) only a very small risk of discarding my working hypothesis $[H]$
prematurely will enable me to demonstrate with assurance $[1-\beta]$ (but
without any unnecessary amount of experimentation) the reality of the influences which I
suspect may be present [$K$].
\end{quote}

In this view, every observation comes with a cost and a major goal of the statistical design of experiments is to make just enough observations in order to convincingly demonstrate a certain effect - $n$ is just as large as necessary, not as large as possible. For example, upon designing a clinical trial, it is now mandatory to calculate the number of patients necessary, given $\alpha$, $\beta$, and $\eta$. More generally speaking, this way to proceed can be extended to an ``a priori power analysis'' \citep{co88, el10}, also well known to psychologists.


\section{Some consequences}\label{standard}

\subsection{The standard style of inference}

Suppose there is an effect $\eta$ of a certain size, and the sample size $n$ is fixed. Then the investigation hinges strongly on the
asymmetry between $\alpha$ and $\beta$, being treated differently. \citet{co66}, p. 21, wasn't the only one to question this choice:
\begin{quote} It is clear that the
entire basis for sequential analysis [and much of received testing theory] depends upon nothing more
profound than a preference for minimizing $\beta$ for given $\alpha$
rather than minimizing their linear combination. Rarely has so
mighty a structure and one so surprising to scientific common sense,
rested on so frail a distinction and so delicate a
preference.\end{quote}
In practice, researchers did not use the additional degree of freedom introduced by \citet{ne33} either. Despite their and Fisher's advice, rather coarse standards such as $\alpha=0.05$, or Cohen's (1988, 1992) classification of effects (small, medium, large) caught on, until testing became a ``ritual'' \citep{gi04b}.

 With all parameters set in advance, a test is indeed a strict decision procedure, and ``the basic objection to this program is that it is too rigid$\ldots$'' (\citet{le93}, p. 70).
 In fact, it is well known that the procedure is so tight that it cannot be extended at all:
\begin{quote} An experimenter, having made $n$
observations in the expectation that they would permit the rejection of a
particular hypothesis, at some predesignated significance level, say $.05$,
finds that he has not quite attained his critical level. He still believes that
the hypothesis is false and asks how many more observations would be required to
have reasonable certainty of rejecting the hypothesis $[\ldots]$\footnote{See also \citet{ro97}, p. 111}

Under these circumstances it is evident that there is no amount of
additional information, no matter how large, which would permit rejection at the
$.05$ level. If the hypothesis being tested is true, there is a $.05$ of its
having been rejected after the first round of observations. To this chance must
be added the probability of rejecting after the second round, given failure to
reject after the first, and this increases the total chance of erroneous
rejection to above $.05$ $[\ldots]$ Thus no amount of additional evidence can be
collected which would provide evidence against the hypothesis equivalent to
rejection at the $P=0.05$ level $[\ldots]$
(\citet{co66}, p. 19)
\end{quote}

In other words: In this perspective, $\alpha$ is a limited, non-renewable resource. ``Once we have spent this error rate, it is gone''
(\citet{tu91}, pp. 104). Thus it has to be used with great care:
$[\ldots]$ a very few prespecified comparisons will be allowed to eat up the
available error rate, and the remaining comparisons have the logical status of
hints, no matter what statistical techniques may be used to study them.
(\citet{tu91}, pp. 104)

In order to avoid an ``inflation'' of error, it seems wise to distribute the error rate of $5\%$ say, among all tests planned. The standard technique is to adjust $\alpha$, a priori, by some scheme taking the whole family of tests into account. \citet{sa85}\label{sals1}, p. 221, reports the consequences of such a consistent attitude:

\begin{quote}
Finally, we should consider the subclass of practitioners who are
`more holy than the Pope,' so to speak. To these practitioners, the
whole purpose of the religion\label{religion} of Statistics is to
maintain the sanctity of the alpha level (which is another name for
0.05). No activity that appears to involve looking at data for
sensible combinations of interesting effects\label{effekt1} is
allowed. It is forbidden, in fact, to do anything more than to
compute the $p$ value using a method determined in advance of the
experiment\label{exp7} and fully documented at that time.

\end{quote}

Note also that if only a small proportion of $\alpha$ is spent in every test, the overall procedure becomes very conservative: In the Neyman-Pearson framework, a very small $\alpha$ corresponds to an inflation of $\beta$ and thus deteriorating power $1-\beta$. Since research in the social sciences is generally plagued by low power, this attitude makes it even more difficult to detect effects.
\citet{el10}, p. 79, concludes:
\begin{quote}
  Instead of dealing with the very credible threat of Type II errors, researchers have been imposing increasingly stringent controls to deal with the relatively unlikely threat of Type I errors \citep{sc92}. In view of these trade-offs, adjusting alpha may be a bit like spending \$1,000 to buy insurance for a \$500 watch.
\end{quote}

\citet{ro91}, p. 57, states another way to deal with the problem described by Tukey. Instead of lowering $\alpha$ for each test, one simply restricts the number of planned tests:
\begin{quote}
$\ldots$do not
allow those who are conducting the trial to look at the results as they
accumulate. That is, $[\ldots]$ conceal the evidence\label{evi16} from the
physician until the trial is completed.\end{quote}

Altogether, the Neyman-Pearson framework gives some justification for minimizing the amount of information collected, and the number of looks at the data. This fits well with Popper's rationalistic view, who always emphasized the role of theory and deduction in the guise of falsification, downplaying the role of data, and rejecting induction firmly \citep{po59, po83}. However, scientific common sense and practice rather point in the opposite direction: 
If we are to learn from experience, an open-minded attitude and any reasonable analysis, be it hypothesis- or data-driven, should be encouraged. \citet{ke95}\label{keid1}, p. 242, admits that
\begin{quote}
$[\ldots]$ it is indeed unsatisfactory to have to defend, perhaps in
the face of senior, highly qualified substantive
scientists,\label{wiss2} our mainstream statistical thinking which
assumes that you are not supposed to look at the data when searching
for methods of optimal analysis with the purpose of gaining new
knowledge.\end{quote}

\subsection{Confusion}

Since there are several theories (at least two), each of them accompanied by a certain ``logic'', data analysis is a tricky business, and there is also lot of confusion.

In particular, despite their mathematical similarity, data-dependent $p$-values and error levels set in advance are completely different. It is against the grain of the Neyman-Pearson theory to calculate $\alpha$-levels a posteriori (for example, one, two or three stars indicating that some empirical result has been significant at the $0.05$, $0.01$ or the $0.001$-level), to report $p$-values instead of zero-one decisions, or to restrict attention to one hypothesis (typically the null, although two hypotheses might be mentioned).
Nevertheless, practice and textbooks use $p$-values and $\alpha$-levels almost interchangeably, thus creating an ``alphabet soup'' \citep{hu04}.

On a less formal level, there is also much conceptual confusion, (inductive) evidence in Fisher's sense and (deductive) decisions in Neyman's and Pearson's being conflated:  \begin{quote}
This hybrid is essentially Fisherian\label{fis21} in its logic, but
it plays lip service to the Neyman-Pearson theory of testing
$[\ldots]$ Some researchers do use the Neyman-Pearson theory of
testing in a pure form, but they constitute a small minority
$[\ldots]$ Regardless of their terminology and verbal allegiance,
most researchers in the fields mentioned above use and/or accept as
valid a pattern of inductive reasoning that is characteristic for
the Fisherian\label{fis22} test of significance. (\citet{sp74}, p. 211)
\end{quote}

It is a crucial ingredient of the standard Neyman-Pearson theory to treat the hypotheses asymmetrically. Typically, the null hypothesis represents the idea that pure chance produced the data at hand, whereas its alternative claims that an interesting substantial effect has left its traces in the data. Obviously, any ``logic of empirical science'' demands that the more data there is, the more difficult it should be for a substantial hypothesis to succeed: ``$\ldots$ in physics and the related disciplines the parent theory is subjected to ever more critical examination as measurement techniques, in their broadest sense, improve. That is, as power increases the `observational hurdle' that the theory must clear becomes greater.'' (\citet{oa86}, pp. 40)

In other words, as information accrues, it becomes easier to detect if the data deviate from a particular hypothesis. For example, suppose your hypothesis (derived from basic theory) claims that about $6.6 \cdot 10^{10}$ neutrinos should hit the surface of the earth per second and $cm^2$. Then measurements should confirm this guess, i.e., the number of neutrinos actually counted should be close to $6.6 \cdot 10^{10}/s\cdot{cm}^2$. In the jargon of statistical tests this means that ``$[\ldots]$ in the physical sciences the substantive theory is associated with the null hypothesis and to the extent that it defies rejection it commands respect'' (cf. \citet{oa86}, p. 41. For a contemporary example see \citet{dy14}.)

However, ``the opposite is the case in the social and behavioural sciences $[\ldots]$ In psychology and the social sciences the substantive theory is associated with the alternative hypothesis and is corroborated as the null hypothesis is rejected. In this sense the observational hurdle which the theory must clear is lowered as power or experimental precision is increased. This is the great weakness of identifying a theory with the alternative hypothesis, to defend such a practice $[\ldots]$ is a nonsense'' (\citet{oa86}, pp. 40).
\begin{quote} Putting it crudely, if you have enough
cases and your measures are not totally unreliable, the null hypothesis will
always be falsified,\label{fals2} {\it regardless of the truth of the
substantive theory} (\citet{me78}, p. 822, emphasis in the original).
\end{quote}

Perhaps it is quite telling that, although this phenomenon was described by an eminent psychophysicist 50 years ago \citep{me67}, and has been decried many times ever since (e.g., \citet{me90, me97,ge13,bo14}), this kind of ``mindless statistics'' \citep{gi04a} has thrived \citep{hu00}. Its ``career'' is quite similar and related to that of $p$-values which, despite their major shortcomings, have also become standard in many sciences, psychology included.

\section{The scientific style of inference}

Scientists, statisticians and philosophers have written much about this state of affairs (e.g., \citet{ne55, ne61, go88, ba99, di11, cu14, sp14, ha16}). Instead of adding another opinion, it may be wiser to go back to the original issue. Since Fisher's elementary setting is too coarse and leads immediately to almost inextricable problems, there is a consensus that two hypotheses should be considered. Alas, the former section shows that Neyman's and Pearson's treatment has led to disappointment. This is quite astounding since the basic problem is rather elementary, and one thus expects an elegant, satisfactory answer.

Looking at the models from a mathematical point of view, the $P$ integral springs to mind. Introduced by Fisher - faute de mieux - it is given the leading part in the standard two-hypotheses setting, and is at the root of all subsequent trouble. More precisely: To keep up the basic distinction between the observed and the unobserved, one has to stick to a strict prior viewpoint. Since this is hardly possible and has curious consequences, it is no coincidence that Neyman's and Pearson's stance has merged with Fisher's position (and later ideas), almost inevitably creating confusion and endless discussion.

Yet, despite all scientific and philosophical turmoil, it is important to see that the origin is just the particular mathematical treatment of the basic problem:
 \begin{enumerate}
   \item Considering intervals like $P_H(X \le x)$ instead of point probabilities,
       \item dealing with the hypotheses in an asymmetric manner, and
   \item putting all parameters constituting a standard test on a par
 \end{enumerate}
makes testing more complicated than it needed to have been:

 a) Power analysis hinges on the idea that the level of significance $\alpha$, power $1-\beta$, effect size $\eta$ and sample size $n$ ``$\ldots$ are so related that any one of them is a function of the other three, which means that when any three of them are fixed, the fourth is completely determined'' (\citet{co88}, p. 14). However, consistently, given $\alpha, \beta$ and $\eta$, this line of thought also supposes that a small sample is optimum \citep{pe55}. More importantly,  formulae such as (\ref{total}) indicate that $\alpha$ and $\beta$ had better depend on $n$. In particular, $\alpha(n)$ should be a decreasing function in $n$ (see, e.g., section \ref{object}, \citet{li57}, and \citet{na16}).

b) The ``scientific'' power of a certain study is not $1-\beta$, but its contribution to a series of experiments, all investigating the same phenomenon \citep{ot96}.
To this end, the effect size $\eta$ is much more important:
\begin{quote}
$\ldots$the emphasis on significance levels tends to obscure a fundamental
distinction between the size of an effect and its statistical significance.
Regardless of sample size, the size of an effect in one study is a reasonable
estimate of the size of an effect in replication (\citet{tv71}, p. 110).\footnote{\citet{gu85}, pp. 3, adds: ``The
emphasis on statistical significance over scientific significance in
education and research represents a corrupt form of the scientific
method$\ldots$''}
\end{quote}

c) Orthodox theory was mainly developed in a time when observations were ``expensive'', i.e., when $n$ was almost always small and typically also fixed. Yet, since information accrues with data, the overall attitude toward $n$ should be quite the opposite:
\begin{quote}
There are no inferential grounds
whatsoever for preferring a small sample $[\ldots]$ the larger the sample the
better $[\ldots]$ The larger the sample size the more stable the estimate of
effect size; the better the information, the sounder the basis from which to
make a decision $[\ldots]$ (\citet{oa86}, pp. 29, 32)
\end{quote}

In everyday life, this often means collecting data until the evidence has built up sufficiently:
\begin{quote}
An experiment involving an image-producing apparatus often
ends appropriately with a `golden event', that is, a picture or image of
something whose existence has been conjectured, but possibly questioned. An
experiment involving a counting apparatus often ends appropriately when a
decision based on some probability model suggests that enough counts have been
taken for some purpose. (\citet{ac89}, p. 189)\footnote{If some insight thus occurs all of a sudden, the crucial last step, with a wink, has been called the {\it interocular traumatic test}: ``You know what the data mean when the conclusion hits you between the eyes'' (\citet{ed63}, also see \citet{bo14}).}
\end{quote}

Altogether, $n$ and $\eta$ seem to be much more important than $\alpha$ and $\beta$. It is also no coincidence that any philosophy \citep{ne77, ma96} based on a suboptimal formal treatment yields results that are at variance with common sense.

The following table summarizes main findings:

\vspace{1ex}
\begin{tabular}{|l|l|l|l|l|}
  \hline
  Approach & Probability model & Math. Treatment & Criticism \\\hline
  Fisher & Observation $x$, and  & simple: 1 hypothesis & underparameterized \\
  & region, e.g., $P(X \ge x)$ & $p$-value & \citet{je39} \\\hline
  Neyman  & Partition: $P_H(X < m)$ & 2 hypotheses $H, K$  & strict prior view \\
  \& Pearson & versus $P_K(X \ge m)$. & $\alpha, \beta, \eta$ and $n$, with & treat. not elegant \\
 ``mainstream''  & Obs. $x \rightarrow$ decision  & emphasis on $\alpha, \beta$ & wrong emphasis \\\hline
  Likelihood & Obs. $x$, and ratio & straightforward:  & posterior view \\
   & $P_K(X = x)$ & 2 hypotheses, and &  \\
   & $\overline{P_H(X = x)}$ & emphasis on $\eta, n$ &  \\\hline
  Bayes & Likelihood ratio & 2 hypotheses  & posterior view \\
   & and prior probs. & evidence {\it and / vs.} & prior probabilities \\
   & $\pi_H, \pi_K$ & prior information & overparameterized \\\hline
\end{tabular}

\section{Testing need not be complicated}\label{easy}

Due to the law of large numbers, testing is an easy problem: If $n$ is not too small, the empirical distribution of the data $P_{X^n}$ is (in any reasonable sense) close to the true distribution $P_H$. The test of one hypothesis gives a formalized answer to the simple question: Is the data I have observed compatible with my hypothesis? If there are two or several hypotheses, the question becomes: Given my set of data, which hypothesis should I choose?
Qualitatively speaking, it is reasonable to choose the hypothesis which is closest to the data, and to reject a hypothesis if the data is ``far away'' from $P_H$. (A formal treatment is given below.)

Given this, the standpoint taken by \citet{ne33}, p. 74, looks rather surprising:
\begin{quote}
 If $x$ is a continuous variable $\ldots$ then any value of $x$ is a
singularity of relative probability equal to zero. We are inclined to
think that as far as a particular hypothesis is concerned, no test based upon a
theory of probability (Footnote: cases will
of course, arise where the verdict of a test is based on certainty$\ldots$)
can by itself provide any valuable evidence of the truth or falsehood of that
hypothesis.
\end{quote}

In the light of the above discussion, these statements - still very popular today - permute rule and exception. They are much too pessimistic, since, owing to the (very) general convergence results, no matter whether the variables are discrete or continuous, given enough observations, $H$ and $K$ can be distinguished with hardly any doubt left. For example, just a few throws suffice to decide between a cube with the numbers $\{0,\ldots,5\}$, and a cube with the numbers $\{1,\ldots,6\}$. More generally speaking, if the support of $H$ and $K$ is not the same (i.e., if there exists some $x$ such that $P_H(x)=0$ and $P_K(x)>0$, or vice versa), one is able to discriminate deterministically between the hypotheses after just a finite number of observations.

Of course, for any continuous random variable $X$, and any realization $x$, $P(X=x)=0$. Therefore, given two such hypotheses, one has to consider their densities, $f_H(x)$ and $f_K(x)$ say. In the case of the normal family (and many others), the support of any two densities coincides. Thus, rather trivially, no matter which $x$ is observed, one cannot decide for sure if $H$ or $K$ is the case. However, the ratio $f_K(x) / f_H(x)$ gives valuable evidence and much more so will $r_n(x_1,\ldots,x_n)$ if $n$ is not too small. Asymptotically, any doubt vanishes completely.

A contemporary treatment, focussing on information and (generalized) distance of distributions, is as follows.
Suppose there are two hypotheses $H,K$, represented by their distributions $P_H,P_K$. Define their KL-divergence \citep{ku51}:
$$
D(P_K || P_H) = \sum_{x \in {\cal X}} P_K(x) \log \frac{P_K(x)}{P_H(x)},
$$
where $D(P_K || P_H) < \infty$, and, for the sake of mathematical simplicity, ${\cal X}$ is a finite set. $D(P_K || P_H)$ may be interpreted as a ``generalized distance'' between distributions; since its introduction, it has become a core concept of information theory and beyond (e.g., see \citet{co06}, pp. 377, and their pointers to the literature).

The key result, connecting the (log) likelihood ratio and KL-divergence is
\begin{equation}\label{likeKL}
 \log \frac{P_K (x_1,\ldots,x_n)}{P_H(x_1,\ldots,x_n)}  = \sum_i \log \frac{P_K(x_i)}{P_H(x_i)}= n \left(  D(P_{X^n} || P_H) - D(P_{X^n} || P_K) \right),
\end{equation}
where $P_{X^n}$ is the empirical distribution of the data.

In the most complete (i.e., Bayesian) setting, $H$ and $K$ are endowed with prior probabilities, $\pi_H$ and $\pi_K$, respectively. Given an iid sample $x_1,\ldots,x_n$ from either $P_H$ or $P_K$, let $A_n \subseteq {\cal X}_n$ be the acceptance region for $H$, depending on $n$. Thus one obtains the error probabilities $\alpha_n = P_H ({\bar A}_n)$ and $\beta_n=P_K(A_n)$, where ${\bar A}_n$ denotes the complement of $A_n$, i.e., the acceptance region of $K$.
Finally, it is straightforward to minimize the total probability of error $P_e(n)=\pi_H \alpha_n + \pi_K \beta_n$.

Thus it turns out (\citet{co06}, p. 388) that ``the optimum decision rule is to choose the hypothesis with the maximum a posteriori probability,'' which means to choose $K$ if
$\pi_K P_K(X_1,\ldots,X_n) > \pi_H P_H(X_1,\ldots,X_n)$, and $H$, if the inequality is in the other direction. Equivalently, the best strategy is a decision in favour of $K$ if
$$
\log \frac{\pi_K}{\pi_H}+ \sum_i \log \frac{P_K(X_i)}{P_H(X_i)} >0 ,
$$
and in favour of $H$ otherwise. Because of (\ref{likeKL}), the latter inequality is tantamount to a decision in favour of $K$ if and only if
$$ \log (1/\pi_H)+n D(P_{X^n} || P_H) > \log (1/\pi_K) +n D(P_{X^n} || P_K).
$$
Without prior probabilities, it is reasonable to decide in favour of $K$ if the empirical distribution is ``closer'' to $P_K$, i.e., if $D(P_{X^n} || P_H) > D(P_{X^n} || P_K) .$
More generally speaking, because of $(\ref{likeKL})$, a decision in favour of $K$ if $P_K(x_1,\ldots,x_n) / P_H(x_1,\ldots,x_n) \ge s$, i.e., if the likelihood ratio exceeds a certain threshold ($s \ge 1$), is equivalent to a decision in favour of $K$ if $D(P_{X^n} || P_H) - \frac{1}{n} \log s \ge D(P_{X^n} || P_K) $. In other words, the likelihood ratio test advises choosing $K$ if the divergence $D(P_{X^n} || P_K)$ is smaller than $D(P_{X^n} || P_H)$ minus the asymptotically vanishing ``safety margin'' $(\log s)/n \ge 0$.
 Moreover, if $K$ is true,
$$
\lim_{n \rightarrow \infty} \frac{1}{n} \log \frac{P_K(X_1,\ldots,X_n)}{P_H(X_1,\ldots,X_n)} \rightarrow D(P_K || P_H) \;\;\; \mbox{in probability}.
$$

     The test closest to Fisher's original idea is Hoeffding's ``universal test'', which merely compares the data with a fixed hypothesis. It decides in favour of $H$ if the empirical distribution is within a certain acceptance region $A_n$ about $P_H$, i.e., if $$D(P_{X^n} || P_H) \le c_n.$$ Because of the law of large numbers, the sequence $c_n$ decreases rapidly with increasing $n$. (For details see \citet{ho65}, theorems 7.1 and 5.1.)

\section{Discussion and conclusions}

At first glance, it seems to be a drawback that KL-divergence is not a proper metric. In particular, given two distributions $P_H$ and $P_K$, in general, $D(P_H || P_K) \neq D(P_K ||P_H)$. However, in the case of data and hypotheses this is an advantage, since there is a striking asymmetry between moving from specific observations to general laws (induction) and the opposite direction (deduction).

Hoeffding's test starts with a hypothesis $H$ and asks if the data lies within a circle of radius $c_n$ about $P_H$. An even more straightforward way to proceed would be to start with data ${\bf x}_n = (x_1,\ldots,x_n)$ and ask if $P_H$ lies within a circle of radius $c_n^{'}$ about the empirical distribution. There is a real difference: Hoeffeding looks for data compatible with some conjectured hypothesis, whereas the second approach conditions on the data and looks for hypotheses that are compatible with the observations.

At least from a mathematical point of view, asymptotically, these preferences do not matter, since, if $H$ is the true hypothesis, by the law of large numbers $P_{X^n}(a) \rightarrow P_H(a)$ for every $a \in  {\cal X}$ in probability (and almost surely). Thus, for every $a$ with $P_H(a)>0$,
$$\lim_{n \rightarrow \infty}  \frac{P_{X^n}(a)}{ P_H(a)}=\lim_{n \rightarrow \infty} \frac{P_{H}(a)}{P_{X^n}(a)}=1$$ which implies $\lim_{n \rightarrow \infty} D(P_{X^n} || P_H) = D(P_H || P_{X^n})=0$ with probability one.

In total generality, i.e., in philosophy, deduction is regarded as rather unproblematic. However, the problem of induction has haunted statistics, philosophy, and perhaps also the sciences at least since David Hume's time \citep{ho03}.
 A standard statistical test is a particularly simple model to study these matters - a ``test bed'' if you allow the play on words. Any such test considers hypotheses (typically two), collects an iid sample $x_1,\ldots,x_n$, and finally decides in favour of or against a hypothesis. Schematically:
$$
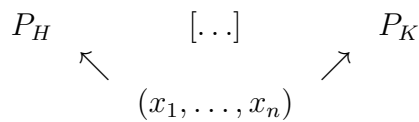
\begin{array}{ccccc}
  P_H  & & [\ldots] && P_K \\
 & \nwarrow  && {\bf \nearrow } & \\
 & & {(x_1,\ldots,x_n)} & &
\end{array}
$$

Philosophers named this kind of reasoning ``inference to the best explanation'' \citep{li04}, but also leading statisticians have always been well aware of the basic issue involved. While \citet{fi35b, fi55} and his school thought in terms of inductive inference, \citet{ne77} sided with the deductive line of argument.
Considering a single experiment, Fisher thus calculates the $p$-value, expressing the {\it evidence} in the data against a hypothesis. Starting with hypotheses, instead,  Neyman and Pearson focus on {\it probabilities of error} and how to control them.\footnote{Quite similarly, Bayesians focus on the data at hand, whereas orthodox theory is much more concerned with the process producing the data.}

In the end, the strong link between Neyman's Frequentist school and Popper's critical rationalism strengthened both points of view, with the consequence that their positions succeeded after the death of their major opponents (Fisher died in 1962, and Carnap in 1970).
In particular, induction was banned, and {\it mathematical} statistics superseded semantic reasoning to an extent that even analyzing given sets of data became suspicious.

Since the 1970s, many statisticians, scientists and philosophers have worked on overcoming this distorted view (e.g., \citet{tu77, he85, gh88, sc92, sc96, be03, ja03, he05, ho06, ri07, hu09, pe09, el10, bo14}). Perhaps since the ``big questions'' thus demanded much attention, rather elementary facts like those pointed out in this contribution could be overlooked easily.

With respect to testing, it is most significant that due to the law of large numbers, the distance between sample and population shrinks (quickly) when $n$ gets larger. This basic insight makes testing an easy problem: Given enough data - and thus information - the true distribution comes into focus almost inevitably. Therefore, mathematically, all approaches based on the straightforward ratio $P_K(X=x) / P_H(X=x)$ lead to unequivocal and strong convergence results. In other words, sufficiently precise and distinct hypotheses can be tested efficiently, at least, if the hypotheses are treated in a rather symmetric way \citep{ro97, ro07}. 

\vspace{1ex}
On a larger scale, testing not only allows for an elegant treatment, but putting information first - e.g., formalized with the help of KL-divergence - gives sound answers to quite a few questions. For example, if there is an uncountable number of hypotheses, e.g., a parameterized family of distributions $P_\theta(x)$, Fisher's likelihood function $L_{\bf x}(\theta)$ provides the key to an elegant solution, which can be extended to an enormously general and powerful information-oriented approach that is perfectly compatible with scientific common sense (e.g., \citet{al97}, \citet{bu02}, \citet{li08}).

Finally, it should be mentioned that particular formal treatments are associated with certain schools of thought - and it is rather the detailed treatment that triggers the overall attitude than vice versa (methods first, philosophy second). Therefore, quite straightforwardly, an elegant mathematical treatment comes with a ``moderate'' and reasonable standpoint, whereas questionable decisions lead to rather ``extremist'' points of view. The above example demonstrates that it may take decades - filled with excessive discussions ranging from formal minutiae to philosophical principles - to overcome a popular, yet distorted, paradigm.

 \vspace{5ex}

\end{document}